\documentclass[11pt,letterpaper]{article}
\usepackage[english]{babel}
\usepackage{amsmath}
\usepackage{amsfonts}
\usepackage{verbatim}
\usepackage{fancyhdr}
\usepackage{fancybox}
\usepackage{graphicx}
\usepackage{amssymb}
\usepackage{fullpage}

\usepackage[T1]{fontenc}
\usepackage[utf8]{inputenc}
\usepackage{authblk}

\usepackage{threeparttable}

\usepackage{amsmath,bm}

\usepackage{ulem}

\newenvironment{hangref}
{\begin{list}{}{\setlength{\itemsep}{4pt} \setlength{\parsep}{0pt}\setlength{\leftmargin}{+\parindent} \setlength{\itemindent}{-\parindent}}}{\end{list}}

\begin{document}



\title{Price Competition with Geometric Brownian motion in Exchange Rate Uncertainty }

\author{Murat Erkoc \\
Department of Industrial Engineering\\
University of Miami, Coral Gables, FL, USA\\

\and Huaqing Wang \\
Department of Management \\
Emporia State University, Emporia, Kansas, USA \\

\and   Anas Ahmed \\
Department of Industrial Engineering\\
King Abdulaziz University, Jeddah, Kingdom of Saudi Arabia \\

}

\renewcommand\Authands{ and }

\maketitle

\linespread{1.35}\selectfont




\maketitle

\linespread{1.35}\selectfont

\begin{abstract}

\noindent \small{We analyze an operational policy for a multinational manufacturer to hedge against exchange rate uncertainties and competition. We consider a single product and single period. Because of long-lead times, the capacity investment must done before the selling season begins when the exchange rate between the two countries is uncertain. we consider a duopoly competition in the foreign country. We model the exchange rate as a random variable. 
We investigate the impact of competition and exchange rate on optimal capacities and optimal prices. We show how competition can impact the decision of the home manufacturer to enter the foreign market. }\\

\end{abstract}

\noindent \rule{\textwidth}{1.5pt}


\section{Introduction}
\linespread{1.5}\selectfont

\normalsize{Exchange rate is one of the important key decisions for a multinational manufacturer to introduce its products to international markets in order to increase its wealth and gain markets share. During the past 20 years, firms are expanding their product and services to foreign market, Lowe et al. (2002). Therefore, if the manufacturer made a decision to enter the foreign market, based on the realization of a favorable exchange rate, it must coordinate its supply chains (supply/manufacture/distribute) properly to maximize its net profit. To achieve this goal, the manufacture has to allocate its capacity properly and flex its prices accordingly. Unfavorable exchange rate will force the manufacturer to not enter the international markets, financial losses due to poor foreign currency exchange. For example, a British company called Lakers Airlines offered low cost tickets for the masses (Jet Blue is an example of current companies that uses their technique) and it was successful, however the company collapsed in 1982 because of the exchange rate. The demand was high and the company was paying for aircraft in strong dollar and earning their profit in weak sterling (Anderson 1997). }

Mainly in this paper, the manufacturer produce its product in the home country and sales occur in both markets, however we will consider the case when the manufacture invests capacity in the foreign country and this capacity will be dedicated only to the foreign demand. The goal of the manufacture is to maximize its net profit which is equal to the difference between discounted expected profit from sales and capacity cost investment.

In this paper, we consider capacity as all the resource available for the manufacture to produce, such resources are technology, funds, man power, logistics, raw materials, and offices. By using multi-period setting, price sensitive demand, and modeling the exchange rate as a Geometric Brownian Motion, we investigate the effect of exchange rate parameters, specifically drift, volatility, and initial exchange rate, on capacity allocation and prices in the two scenarios. Also, we show the cases: when it's optimal for the global manufacturer to serve both markets, deny service to one market, and invest capacity in the foreign market. Finally, we show the effect of the planing horizon on optimal capacity, and investing capacity at the foreign market.

This paper is organized as follows. In the preceding section, we present the literature review. In section 3, we present our basic settings, assumptions, and nomenclature. In section 4, we present the Early Commitment to Price and Quantity Model where the manufacturer commits to prices and quantities at the beginning of the planning horizon. In section 5, we present the Postponement of Prices and Quantities Model where the manufacturer flexes its prices and quantities from period to period. For each model we discuses the allocation scenarios under capacity investment at home and in both countries. In section 6, we will have a summary and conclusion.

\section{Literature Review}
\linespread{1.5}\selectfont

\normalsize{Multinational firms are facing daily challenges such as demand volatility, governmental regulations, distribution constraints, and exchange rate uncertainties. The area if this proposal is to tackle the exchange rate uncertainties. Exchange rate fluctuations from 1\% a day to 20 \% a year is common and this changes could affect capacity and cost, Dornier (1998), which will ultimately affect profit. For a global firm to continue operating internationally and stay in business, the firm must be flexible and proactive in responding to uncertain shocks in exchange rate movements and adjust its operations based on unexpected movement by exchange rate. Therefore, multinational firms incorporate multiple risk-management techniques to handle exchange rate uncertainty. }

Two types of techniques have been developed by researches to tackle such a challenge: financial hedging and operational hedging. The multinational firm has to apply one of the two methods or combine both methods, an e.g. of combining both strategies done by Ding, Dong, and Kouvelis (2007), to gain market share and increase the firm's wealth and prosperity. Before operational hedging, there was financial hedging where the multinational firms buy option contracts; an example of financial hedging is currency option: a contract that grants the multinational firm the right to buy or sell currency at a specified exchange rate during a specified period of time. The multinational firm can opt out of the contract if the future exchange rate is not favorable. According to O'Brien (1996), currency option is used regularly by firms to protect itself against exchange rate volatility. Another example that used frequently is forward exchange contract: an agreement between two parties to exchange a specified amount of one currency for another currency at a specified foreign exchange rate on a future date. Another advantage of using this method, besides hedging against hedging against unfavorable exchange rate fluctuations, is cost and budget are accurate. However, there are critical disadvantages that could hit the firm with financial losses. Huchzermeier and Cohen (1996) assert that the firm could loss profit because the firm will not catch in the upside of exchange rate volatility and it could be risky for a firm to enter foreign market. They also mentioned that if the firm is not hitting the targeting sales and the exchange rate is weak, then the result of implementing the forward contract will be financial loss for the multinational firm. Any cancellation of the contract will result in a financial loss for the multinational firm.

This work focuses on operational hedging; where we propose operational hedging mechanism for the firm to implement to hedge against exchange rate fluctuations. In this dissertation, we do not implement any financial hedging on our model for the reasons mentioned above. Hence, operational hedging could also be used to hedge against demand uncertainty (see Mieghem and Dada 1999).

Operational hedging strategies is clearly defined and illustrated in Cohen and Huchzermeier (1999), Cohen and Mallik (1997), and Kouvelis (1999). Ding et al. (2007) defined operational hedging strategies as "real compound options that are exercised in response to the demand, price, and exchange-rate contingencies faced by the firm in a global supply chain context." Such operational techniques (real options) can be found in the literature. Some work include postponing prices (Kazaz, Dada, and Moskowitz 2005), switching production between countries (Dasu and Li 1997), choosing the optimal path in the supply chain network (Huchzermeier and Cohen 1996), reserving extra capacity (Cohen and Huchzermeier 1999), and producing less than the total demand before the selling period (Kazaz, Dada, and Moskowitz 2005).

Operational flexibility consists of the above real options. According to Cohen and Huchzermeier (1999), changes in market conditions, e.g. exchange rate or demands, are anticipated by the firm's operational flexibility. By implementing these options, the multinational firm can take advantages of the market volatility.

{Huchzermeier and Cohen (1996) consider a supply chain consisting of supplying countries, production facilities, and markets. Similar to our model, capacity decision is made before realizing the exchange rate. They developed model to select the optimal supply chain. For example, one optimal supply chain option is to supply the material form home country and manufacture the product at all factories and each factory sells it to its market. They implement a multinomial approach to approximate exchange rate. In their paper, they showed how operational hedging can lower the risk of exchange rate fluctuations. Our work is different, we save money by eliminating switching cost between production locations and we assume a multi-period horizon where the firm can change its prices from period to period, our third model, to hedge against exchange rate.

Kogut and Kulatilaka (1994) implemented shifting production between factories located in different countries, depending on exchange rate movement, as a mean of operational hedging. Our work differs in the sense that we do not shift production between countries, we use pricing as a hedging mechanism.

Kouvelis et al. (2001) introduce a new and interesting kind of shifting; shifting between strategies. Moreover, they proposed three types of strategies: exporting where capacity is at home, joint venture with a firm at the foreign market, and investing capacity at the foreign market. Exchange rate volatility can affect the shifting between those strategies. For example, increasing volatility would force the firm to implement the exporting strategy. In their work, they assume that the firm entered the foreign market, our work differs in a sense that we show when it is profitable to enter the foreign market, because it be costly to enter the foreign market under unfavorable exchange rate as we will show in our first model.

Kouvelis and Rosenblatt (2001) show how government subsidies such as: tariffs, financing of facilities and rules of trading can affect global manufacturing and distribution networks. They provided a mixed-integer program to evaluate the affected of those factors. We differ in our analytical approach.

Kazaz et al. (2005) developed a two staged stochastic program. The goal of this program is to hedge against uncertain exchange rate. This program chooses the optimal allocation policy based on the realized exchange rate. In their model, capacity is built and unites are manufactured before the realization of the exchange rate. Then, the exchange rate is know, the firm will decide if it is going to serve all markets or serve one market and deny the other market. They called the first step ``production hedging'' and the second step ``allocation hedging''. Similar to this proposal, they employed multiple period model. However, they do invest capacity at the foreign country unless it is more profitable, otherwise we assume that capacity investment is made at home. Also, if the manufacturer decided to invest capacity at the foreign market, that capacity will be used only for the foreign demand.

Ding et al. (2007) present a multinational firm that invests capacity in either one of the two markets or one in each market. They assume that the firm will make capacity commitment before the realization of the exchange rate and demand in both markets. In their paper, they used both financial hedging and operational hedging in order for the firm to protect itself against exchange rate uncertainty. The multinational firm will employ financial hedging, at the time of capacity building, by purchasing financial option contracts. This scheme is famous and mentioned by several researchers such as: Mello, Parsons, and Triantis (1995) and Chowdhry, and Howe (1999). The other method is operational hedging where the multinational firm postpones its allocation of capacity to both markets until the demand and exchange rate are known.

Ding et al. (2007) is similar to Kazaz et al. (2005) paper in a sense that both implement a two-stage stochastic program. In the first stage, both of them invest capacity before the realization of the exchange rate and in the second stage they allocate capacity after the realization of the exchange rate. The main difference is that ding et al. (2007) incorporate financial hedging techniques in the fist stage and Kazaz et al. (2005) look at the multi-period planning horizon.

There are two key differences that distinguish our model from Kazaz et al. (2005) and Ding et al. (2007): the modeling of the exchange rate and the demand function. We assume that the exchange rate follows a geometric Brownian motion and, in Kazaz et al. (2005), they assume that exchange rate is modeled as distributed random variables while, in Ding et al. (2007), they assume that exchange rate follows a lognormal distribution. Even though, in Ding et al. (2007) , they show the affect of exchange rate volatility on optimal capacity and expected profit but it lack more parameters such as: exchange rate drift and initial exchange rate which could greatly affect optimal capacity and expected profit. We assume that our demand is to be a price sensitive demand, additive demand, while Ding et al. (2007) and Kazaz et al. (2005) assume demand is a random variable. A minor difference between our work and Ding et al. (2007) is that they do not take into consideration transportation cost; if transportation cost is relatively big, this could affect the allocation of capacity between the two markets.

Aytekin and Birge (2004) favors financial hedging when exchange rate has a small volatility and operational hedging when exchange rate has a strong volatility. They modeled their problem as a continuous time while we used multi-period setting. They looked at three scenarios where production is done at home, production is done at home and foreign country and home production supply the shortage at the foreign market, and production is done in both countries and each production country can serve either market.

Competition has a major impact on pricing. Bitran and Caldentey (2003) claim that selecting a certain pricing policy depends mainly on the price competition among players. Chan et al. (2004) claim that product price and service are the major factors that firms compete on and that should be taken into account by firms when setting the pricing policies. The work done in the area of competition is ample, however most of it are trying to drive the optimal prices and quantities under demand uncertainty for different demand functions. Here an example of some: Bernstein and Federgruen (2004) addresses the problem of two echelon system (one supplier and competing retailers) under demand uncertainty. They derive the Nash equilibrium in terms of price and base-stock level for each retailer. Also, they show the impact of different parameters such as cost, and distribution parameters on optimal price and optimal base-stock level of a retailer.

Mieghem and Dada (1999) shows how price postponement strategy can generate more profit to the firm under competition and demand uncertainty. Anupindi and Jiang (2008) show that flexible firms (price and quantity postponement) can make more profit that inflexible firms (price postponement) under demand uncertainty (multiplicative and additive demand). However, Bernstein and Federgruen (2004), Mieghem and Dada (1999), and Anupindi and Jiang (2008) did not consider the impact of exchange rate uncertainty on capacity investment and price.

This work is unique in a sense that we are combning exchange rate uncertantity and competition and we investigate the impact of them on optimal prices, optimal capacity, and capacity allocations.

\renewcommand{\theequation}{3.\arabic{equation}}
\setcounter{equation}{0}

\section{Basic Settings, Assumption, and Nomenclature}

We consider a multinational firm that produces a single product at home and sell it to domestic and foreign markets. At the foreign market, we consider a foreign manufacturer that manufacturers the same product and sell it to the foreign market only. Therefore, the firm is competing with the foreign manufacturer on market share at the foreign market only. We consider a monopolistic setting at domestic market, and a duopoly setting at the foreign market, Figure \ref{basic model} (Wang 2014). Such examples of duopoly is: Visa Vs Mastercard, Pepsi Vs Coca-Cola, and Intel Vs AMD in the Microprocessor market. 
\begin{figure}[ht]
\begin{center}
  \includegraphics[scale=0.5]{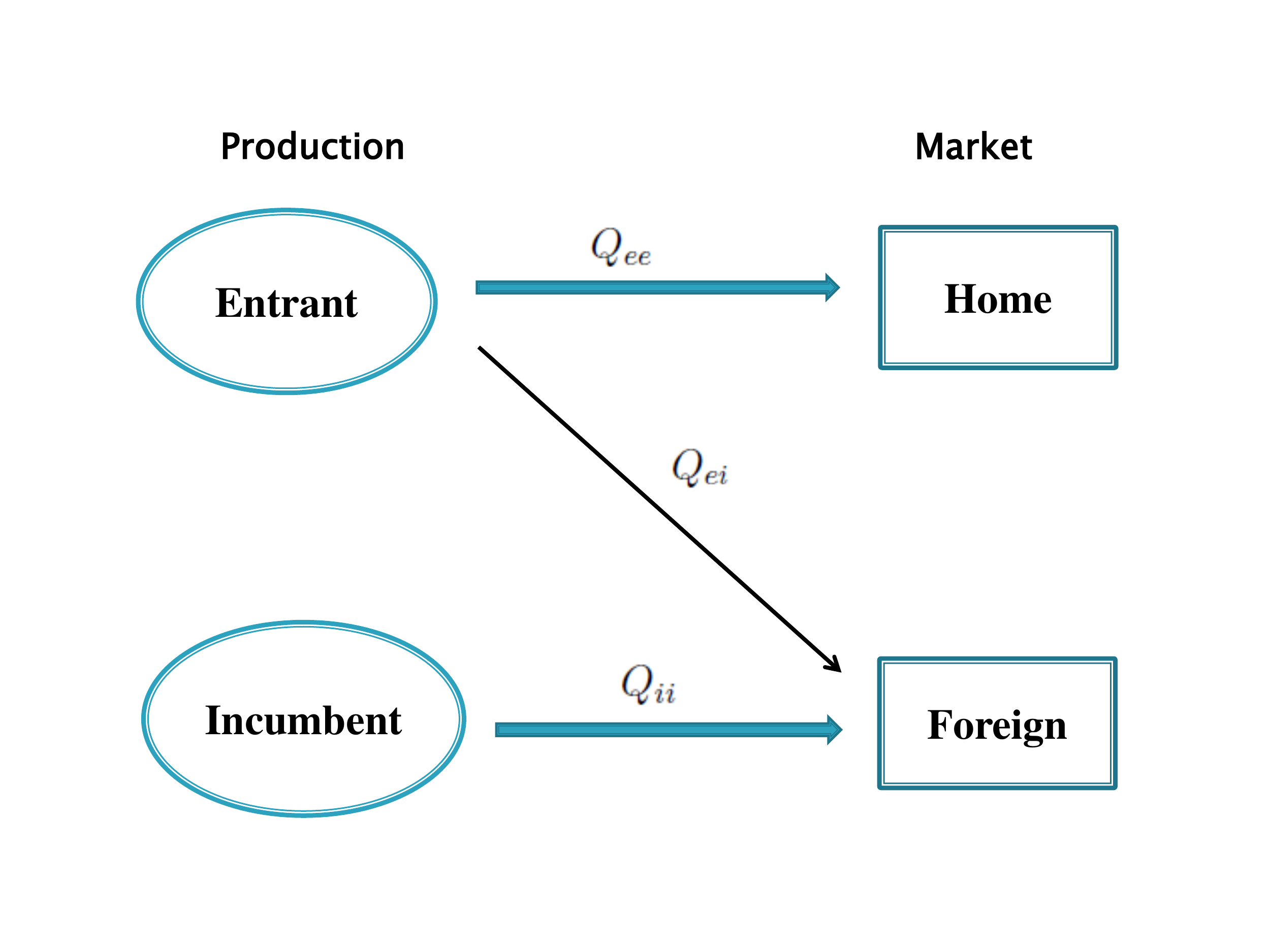} \\
  \caption{The Model}
\label{basic model}
\end{center}
\end{figure} 

\indent We consider a single period where the firm must decide on its initial capacity investment before the selling season,  selling price at each market, and the production allocation across the selling season after the realization of the exchange rate. In this model, the capacity investment is irreversible and must be made in the home country only. The domestic and the foreign markets differ in their demand base and disparity in currencies. In each market demand is assumed to be linearly decreasing in price which is expressed in local currency. The demand base is steady. In this paper, we assume a deterministic demand function of prices and can be "interpreted as expected demand in many applications" (Gallego and Hu 2006). We model the exchange rate as a random variable. It is assumed that the firm does not carry any inventory. The product can be produced and shipped to both markets at the beginning of the selling season. We used the following nomenclator in our analysis:

\noindent {\bf Parameters}

$n$: players index where n = $i$ for the incumbent, and n = $e$ for the
entrant

$j$: market index where $j=i$ for the incumbent's market, and $j$ = $e$ for the
entrant's market


limited capacity

$I_o$: The initial foreign exchange rate expressed in entrant's (home) currency per unit in incumbent's (foreign) currency.

$I$: exchange rate between the currency of entrant's home market and the incumbent's market

$C_i$: unit manufacturing cost at market $\,i$ expressed in market
$i\,$currency

$s$: unit shipment cost to the incumbent market expressed in the entrant's home currency


{\bf Decision Variables } 

$K$: \  capacity invested by the entrant,  


$P_{ij}$: \ unit selling price set by the player $i$ in market $j$

$Q_{ij}$: \ quantity sold by the player $i$ in market $j$ 

\noindent We assume that demand functions are differentiable and satisfy the following conditions:

\begin{equation*}
\frac{\partial D_{ei}}{\partial P_{ei}},\,\frac{\partial D_{ii}}{\partial
P_{ii}}\leq 0  \tag*{$(C1)$}
\end{equation*}
\begin{equation*}
\frac{\partial D_{ei}}{\partial P_{ii}},\,\frac{\partial D_{ii}}{\partial
P_{ei}}\geq 0  \tag*{$(C2)$}
\end{equation*}

Conditions $(C1)\,$and $(C2)\,$are common conditions in the oligopoly pricing literature (see Bernstein and Federgruen (2005) and Gallego and Hu (2006)). $(C1)$ shows that each player's demand decrease in its price. $(C2)\,$indicates that each player's demand increase in the other player's price. We are considering the additive demand function mentioned in Varian (1992), which satisfies $(C1)\,$and $(C2)$:
\begin{equation}
D_{ij}=\alpha_{ij}- P_{ij}+\theta P_{\text{-}ij} \ , 
\end{equation}

\noindent where $D_{ij}$ denotes the demand for player $i\,$ in market $j\,$ while $\alpha_{ij}$ 
represents the demand potential 
for manufacturer $\,i\,$ in market $\,j\,$. 
$\theta\,$ is the ``cross-price effect'' which is symmetric for ``well-behaved consumer demand function''. $\theta$  measures the degree of competition between the two manufacturers (Vives (1999) and Varian (1992)). In this study we focus on non-complementary product (substitutable product). According to Baumann et al. (1998) positive $\theta$ (e.g., an increase in the foreign manufacturer price increases the demand of the firm) means the products of both manufacturer are substitutable and if $\theta$ is a negative value that means that the two products are complements. Therefore, equation (3.1) can be formulated as follows (see Wang et al. (2016))  :
\begin{equation*}
D_{ee}=\alpha_{e} - P_{ee} \tag*{$(3.1A)$} 
\end{equation*}
\begin{equation*}
D_{ei}= (1-\phi) \alpha_{i} - P_{ei}+\theta P_{ii} \tag*{$(3.1B)$} 
\end{equation*}
\begin{equation*}
D_{ii}=  \phi \alpha_{i} -  P_{ii} + \theta P_{ei} \tag*{$(3.1C)$}
\end{equation*}



The total size of the imcumbent manufacturer's market is equal to $\alpha_{i}$, and $\phi $ represents the consumer's preference for its product. 

In our setting, the exchange rate fluctuations across periods are modeled by a Weiner Process, $B(t)$, where $B(t)=\epsilon\sqrt{t}\,.\,$Here, $\epsilon$ is the random error term that follows the Standard Normal distribution. As such, it is assumed that the exchange rate follows a Geometric Brownian motion:

\begin{equation}
dI(t)=\mu I(t)dt+\sigma I(t)dB(t)  \ . \label{brownian}
\end{equation}

where $I(t)\,$is driven by the Ito process. The parameters $\mu$ and $\sigma$ are the mean and the standard deviation of the Normal exchange rate drift. The solution to equation \ref{brownian} is given by Davis (2001) as $I(t)=I_0e^{\left(\left(\mu-\frac{1}{2}\sigma^2\right)t+\sigma B(t)\right)}$. Assuming $B(t)=\epsilon\sqrt{t}$ and \ replacing in \ref{brownian}, we get

\begin{equation}
I(t)=I_0 e^{\left(\left(\mu-\frac{1}{2}\sigma^2\right)t+\sigma\epsilon\sqrt{t} \right)} \ .
\end{equation}

\renewcommand{\theequation}{4.\arabic{equation}}
\setcounter{equation}{0}

\section{The Duopoly  Model}

In this section, we are assuming that both firms make their prices simultaneously at the incumbent's market. We consider two scenarios, 1) both firms has ample (unlimited) capacities,  and 2) the entrant has limited capacity and the incumbent manufacturer has unlimited capacity. We will derive the Nash equilibrium (optimal prices and quantities) for each manufacturer and show the impact of exchange rate and competition parameters on the manufacturer's capacity allocations. 

The entrant manufacturer's objective function is given by
\begin{equation}
\max_{P_{ei}, \, P_{ee} \geq 0}\Pi_{e}=\left(P_{ee} -C_e\right) \left(\alpha_{e} -   P_{ee} \right)+\left(I P_{ei} - C_e - s \right) \left( (1-\phi) \alpha_{i}- P_{ei}+ \theta P_{ii} \right)   \label{entrant profit function}
\end{equation}
\begin{equation}
s.t.\,\, Q_{ee} + Q_{ei} \leq K  \ ,  \label{constraint condition 1}
\end{equation}
where,
\begin{equation*}
Q_{ee} = \alpha_{e} - P_{ee}  \tag*{$(4.1a)$} \,\,\, \text{and} \,\,\, Q_{ei} = (1-\phi) \alpha_{i}- P_{ei}+ \theta P_{ii} \ .
\end{equation*}
The incumbent manufacturer's optimization problem is given by: 
\begin{equation}
\max_{P_{ii} \geq 0} \Pi_{i}=\left(P_{ii}- C_i\right)\left( \phi \alpha_{i} -  P_{ii} +\theta P_{ei}\right) \ .  \label{incumbent profit function} 
\end{equation}



\subsection{Entrant Manufacturer with Unlimited Capacity}

When the entrant's capacity is unlimited, then constraint condition give in (\ref{constraint condition 1}) is relaxed. The entrant manufacturer will set its prices according to the objective function given in (\ref{entrant profit function}).  Solving the problem with respect to $P_{ei}$ and $ P_{ee}$, we get:
\begin{equation}
P^{u}_{ei}(P^{u}_{ii}) =  \frac{(C_e + s) + I ((1-\phi) \alpha_{i} +\theta   P^{u}_{ii})}{2 I} \label{entrant price incumbent un}
\end{equation}
\begin{equation}
P^{u}_{ee} (P^{u}_{ii}) = \frac{\alpha_{e}+ C_e}{2} \label{entrant price own un}
\end{equation}
The incumbent manufacturer will set its prices according to the objective function given in (\ref{incumbent profit function}) and 
we can get  
\begin{equation}
P^{u}_{ii} (P^{u}_{ei}) = \frac{\phi \alpha_{i}+ C_{i}+\theta  P^{u}_{ei}}{2} \ . \label{incumbent price un}
\end{equation}
Simultaneously solving (\ref{entrant price own un}), (\ref{entrant price incumbent un}), and (\ref{incumbent price un}) for $P^{u}_{ee}$, $P^{u}_{ei}$, and $P^{u}_{ii}$, we can get the equilibrium outcomes of $P^{u}_{ee}$, $P^{u}_{ei}$, and $P^{u}_{ii}$ and hence other equilibrium results given in the Appendix.

Substituting $P^{u}_{ee}$ into 
$ D_{ee}=  \alpha_{e} - P^{u}_{ee}  $, we can get the optimal quantities the entrant manufacturer will sell in its own market:
\begin{equation}
Q^{u}_{ee} = \frac{\alpha_{e} -  C_{e}}{2}  = K_Q \ . \label{entrant quantity own unlimited}
\end{equation}
Substituting $P^{u}_{ei}$ and $P^{u}_{ii}$ into $
D_{ei}= (1-\phi) \alpha_{i} -  P_{ei}+\theta P_{ii}  
$, we can get  the optimal quantities the entrant manufacturer will sell in the incumbent market as: 
\begin{equation}
Q^{u}_{ei} = \frac{ I (\alpha_i ( 2  -2  \phi+\theta  \phi )+ C_{i} \theta )  - (C_e+s) (2 -\theta ^2)}{I \left(4  - \theta ^2 \right)} \label{entrant quantity incumbent unlimited}
\end{equation}

\noindent \MakeUppercase{L{\small emma }}4.1: {The firm will export to foreign market if
\begin{equation}
I > \frac{(C_e + s) \left(2 -\theta ^2\right)} {\alpha_i ( 2  -2  \phi+\theta  \phi )+ C_{i} \theta} = I_{z}  \label{iz}
\end{equation} Otherwise, the firm will sell only to domestic market.
}

Equation (\ref{entrant quantity incumbent unlimited})  and  (\ref{iz})  show that it is more attractive to enter into the incumbent's market  as  $C_i$ and $\alpha_{i}$ increase. That is, if the market potential or the production cost of the incumbent market is high, then the entrant manufacturer has more incentive to export their products into the incumbent's market. This result can explain why there are so many manufacturers are pouring their products into the U.S. market since the potential demand in the U.S. is very big and the production cost (including the labor cost) is also very high. \\

If the exchange rate $I(t)$ is low, i.e.,  $ I < I_{z} $, then even the entrant manufacturer has unlimited capacity, it will still not enter into the incumbent's market. Therefore, the entrant manufacturer will sell only to it's own domestic market with profit given by: 
\begin{equation}
\Pi^{EU}_{e} = \frac{(\alpha_e -  C_{e})^2}{4} \ . \label{entrant home profit unlimited}
\end{equation}
Here we slightly abuse the notation by using "$EU$" as the case of "manufacturer selling only to the entrant's own market with unlimited capacity". 
When the exchange rate is enough high, i.e.,  $ I >I_{z}$,  then the entrant manufacturer can export to the incumbent market and compete with the incumbent manufacturer. Substituting $P^{u}_{ei}$ and $P^{u}_{ii}$ into the entrant's objective function, we can get the entrant's profit: 
\begin{equation}
\Pi^{BU}_{e} = \frac{(\alpha_e - C_{e})^2}{4}+\frac{ \left(I (\alpha_i ( 2  -2  \phi+\theta  \phi )+ C_{i} \theta )  - (C_e+s) (2 -\theta ^2) \right)^2}{I \left(4 - \theta ^2 \right)^2} \label{entrant profit unlimited}
\end{equation}
Here we use the notation by using "$Ub$" as the case of "selling in both market with unlimited capacity".  We can see from equation (\ref{entrant profit unlimited}) that  the profit generated from entrant's own market is independent of the price competition in the incumbent's  market and the exchange rate when entrant's capacity is unlimited. The second term is the profit generated from the sales at the incumbent's market and this profit is zero if the exchange rate falls below the threshold value $ I_{z}$. \\

Substituting $P^{u}_{ei}$ and $P^{u}_{ii}$ into the incumbent's objective function, we can get the incumbent's profit function as:
\begin{equation}
\Pi^{u}_i =\left(\frac{\theta  (C_{e}+s) - I (C_{i}(2-\theta ^2) - \alpha_{i} (2-\theta ) \phi -\alpha_{i} \theta )}{(4 - \theta^2) I}\right)^2 \ .
\end{equation}

\noindent \MakeUppercase{C{\small orollary }} 1: {\it The profit for the incumbent firm is monotonically decreasing in the exchange rate when the entrant manufacturer has unlimited capacity.}

This corollary shows how the entrant can benefit from higher exchange rate and as a result, we can derives the next lemma. \\

\noindent \MakeUppercase{L{\small emma }}4.2: {The incumbent manufacturer can not make sales if
\begin{equation}
I > \frac{\theta  (C_e + s)}{C_{i} \left(2 - \theta^2) - \alpha_{i} (\theta  (1- \phi ) + 2  \phi )  \right)}  = I_v   \label{I v unlimited}
\end{equation}
where
\begin{equation}
C_i > \frac{\alpha_{i} (\theta  (1- \phi ) + 2  \phi )}{2 - \theta^2}  = C_v   \label{C v unlimited}
\end{equation}
if $ C_i < C_v$, then the entrant can not force the incumbent manufacturer out of the game. }

The above lemma shows that if the exchange rate is above a certain thresholds, $I_v$, then the incumbent manufacturer will be forced out of its own market by the entrant manufacturer  if and only if $C_i > C_v$. Intuitivley, if the production cost of the incumbent is not too high then it is not possible that the entrant could drive the incumbent market out of the market if the exchange rate satisifies $I  > I_v$. 

\subsection{Entrant Manufacturer with Limited Capacity}

The entrant manufacturer and incumbent face the same objective functions as the case in the unlimited capacity. However, with limited capapcity, the constrait condition (\ref{constraint condition 1}) will be activated.

The entrant manufacturer will set its prices according to the objective function given in (\ref{entrant profit function}).  Using Lagrange relaxation and solving the problem with respect to $P_{ei}$ and $ P_{ee}$, we get:
\begin{equation}
P_{ee}(P_{ii}) = \frac{ I (2 \alpha_{e} +\alpha_{i} - \alpha_{i} \phi  -2 K+ \theta  P_{ii}) + ( \alpha_{e}-  s)}{2  (1 + I)}  \label{entrant price own}
\end{equation}
and
\begin{equation}
P_{ei}(P_{ii}) = \frac{ I  (\alpha_{i} -\alpha_{i} \phi +\theta  P_{ii} ) +   (\alpha_{e} + 2 \alpha_{i} (1-\phi) - 2 K + 2 \theta  P_{ii} +\beta_{ee} s)}{2  (1 + I)} \ . \label{entrant price incumbent}
\end{equation}
The incumbent manufacturer will set its prices according to the objective function given in (\ref{incumbent profit function}) and we can get:
\begin{equation}
P_{ii}(P_{ei}) = \frac{ \phi \alpha_{e} +  C_{i}+\theta  P_{ei}}{2} \ . \label{incumbent price}
\end{equation}
Simultaneously solving (\ref{entrant price own}), (\ref{entrant price incumbent}), and (\ref{incumbent price}) for $P_{ee}$, $P_{ei}$, and $P_{ii}$, we can get the equilibrium prices of $P^{*}_{ee}(K)$, $P^{*}_{ei}(K)$, and $P^{*}_{ii}(K)$,  equilibrium quantity of $Q^{*}_{ee}(K)$, $Q^{*}_{ei}(K)$, and $Q^{*}_{ii}(K)$, and hence both players' profit $\Pi^{*}_{e}(K)$, $\Pi^{*}_{i}(K)$.  
In equilibrium, the entrant manufacturer's profit can be given as:
\begin{equation}
\Pi_{e}^{BC}(K)=\left(P^{*}_{ee}(K) -C_e\right) \left(\alpha_{e} -  P^{*}_{ee}(K) \right)+\left(I P^{*}_{ei}(K) - C_e - s \right) \left( (1-\phi) \alpha_{i}-  P^{*}_{ei}(K)+ \theta P^{*}_{ii}(K) \right)  
\end{equation}
Here, we use the notation "BC" as the entrant selling in both markets with limited capacity. \\

\noindent \MakeUppercase{P{\small roposition}} 4.1: {The firm's capacity is fully utilized if and only if }
\begin{equation}
I \geq \frac{2 (C_{e} + s) (2 -\theta^2)}{ 4 \alpha_{i} - 2 \alpha_{i} (2- \theta) \phi + 2 C_{i} \theta + (4 - \theta ^2) (\alpha_{e} - C_{e} -2 k)} = I_{t}      \label{It}
\end{equation}

\noindent or
\begin{equation}
K < \frac{4 \alpha_{i} - 2 \alpha_{i} (2- \theta) \phi + 2 C_{i} \theta + (4 - \theta ^2) (\alpha_{e} - C_{e})}{2 \left(4 -\theta^2 \right)} = K_{t} \label{Kt} 
\end{equation}

\noindent Otherwise, the firm will sell constrained optimal.

The above proposition shows that the entrant can implement the unconstrained solution and hence does not need to split her capacity across markets if $I(t) < I_{t}$. In contrast, if the inequality in (\ref{It}) does not hold then the capacity is limited. As such, the entrant faces the problem of allocating the capacity between the two markets. Depending on the market potentials and the realized exchange rates the entrant may also choose to sell only in one of the markets. The next proposition indicates that if the exchange rate is significantly high, the entrant manufacturer in fact may opt out of its own market. \\

\noindent \MakeUppercase{P{\small roposition}} 4.2: {In equilibrium, (a) the entrant manufacturer chooses to sell in the domestic market if and only if
\begin{equation}
I < \frac{ (2 - \theta ^2) (\alpha_{e} +  s)}{  (\alpha_{e} (2  (1-\phi) + \theta  \phi )+  C_{i} \theta ) - K \left(4 - \theta ^2\right)} = I_{h}   \label{Ih2} 
\end{equation}
otherwise, the firm sells only in the foreign market; (b)  entrant manufacturer will export its product to the incumbent manufacturer's market if and only if 
\begin{equation}
I > \frac{(2 - \theta ^2) (\alpha_{e}+ s - 2 K)}{ \alpha_{i} (\theta  \phi + 2 (1-\phi))+ C_{i} \theta } = I_{f}  \label{If2}
\end{equation}
That is, the entrant splits the capacity between the domestic and foreign markets only when $I_{f}<I(t)<I_{h}$. }

Note that from (\ref{Ih2})  the entrant manufacturer will always sell to the domestic market (its own market) if 
\begin{equation}
K \geq \frac{ \alpha_{i} (\theta  \phi + 2 (1 - \phi))+ C_{i} \theta}{4  - \theta ^2} = K_{h} \ . 
\end{equation}
When the capacity is unlimited, it is always profitable to sell in the entrant's own market. However when capacity is scarce, as the foreign exchange rate increases, it becomes less appealing to allocate capacity to its own market. 

Proposition 4.2 also shows how competition can increase the appeal to enter the foreign market especially when cost of production ($C_i$),  potential demand alpha ($\alpha_i$) for the incumbent manufacturer, and cross price effect ($\theta$) increase. Proposition 4.2 implies that the optimal policy for the entrant manufacturer at any given period is a ``cherry-picking'' policy when the capacity is scarce. That is, when the exchange rate is significantly high the entrant manufacturer sells only in the incumbent market whereas when it is too low she will sell exclusively in the domestic market. For exchange rates is not too high or too low, the capacity is split between the two markets. \\ 


\noindent \MakeUppercase{L{\small emma }} 4.3: (a) When the entrant manufacturer with limited capacity  sells only in its own market, the entrant's profit is given by 
\begin{equation}
\Pi_e^{EC} = (\alpha_{e} -  C_{e} - K) K \ ;
\end{equation}
(b) when the entrant manufacturer with limited capacity  sells only in the incumbent's market, the entrant's profit is given by 
\begin{equation}
\Pi_e^{IC} = \frac{[I (2 \alpha_{i}  (1- \phi) + \alpha_{i} \theta  \phi +  C_{i} \theta -2  K) - (C_{e}+s) ( 2 - \theta ^2 ) ] K}{2 -\theta ^2} \ . 
\end{equation}
\noindent \MakeUppercase{L{\small emma }} 4.4: The entrant manufacturer's profit function $\Pi^{IC}_{e}(K)$, $\Pi^{BC}_{e}(K)$,  $\Pi^{EC}_{e}(K)$ are concave functions of its capacity $K$.





\noindent \MakeUppercase{L{\small emma }} 4.5: If $\alpha_{e} >  \frac{\alpha_i (2(1-\phi) + \theta \phi) + C_{i} \theta }{4 - \theta ^2} +C_e $ then the entrant manufacturer will always be willing to sell in its own market.

This lemma shows that when the entrant manufacturer's market size is higher than some threshold value then the entrant will never give up its market.


\subsection{The Endogenous Model Optimal Capacity}

\noindent \MakeUppercase{L{\small emma }} 4.6:  Define $K_{Q} = \frac{(\alpha_{e} -  C_{e})}{2} $. (1) If $K_Q < K_{h} < k_{t}$ then $I_{h}>I_{f}$ on $[0, K_Q]$; $I_{h}>I_{f} > I_z $ on $[K_Q, K_h]$; and $I_{t}>I_{z}$ on $[K_h, K_t]$; (2) If $K_{h} < K_Q < k_{t}$ then $I_{h}>I_{f}$ on $[0, K_h]$;  $I_{t}>I_{z}$ on $[K_Q, K_t]$.  \\

\subsubsection{$K_Q < K_{h} < k_{t}$}
Based on Proposition 4.2 and Lemma 4.6, we conclude that at the beginning of the planning horizon the optimal pricing and allocation policy falls in one of the following five scenarios:

\noindent EU:  The capacity is unlimited (unconstrained) and the entrant manufacturer sells only in its own market. \\
BU:  The capacity is  unlimited (unconstrained) and the entrant manufacturer sells in both markets. \\
EC:  The capacity is limited and the entrant manufacturer allocates all of its capacity to its own domestic market.\\
IC:  The capacity is limited and the entrant manufacturer allocates all of its capacity to the incumbent market.\\
BC:  The capacity is limited and the entrant manufacturer allocates all of its capacity between the two markets. \\

\begin{figure}[ht]
\begin{center}
  \includegraphics[scale=0.5]{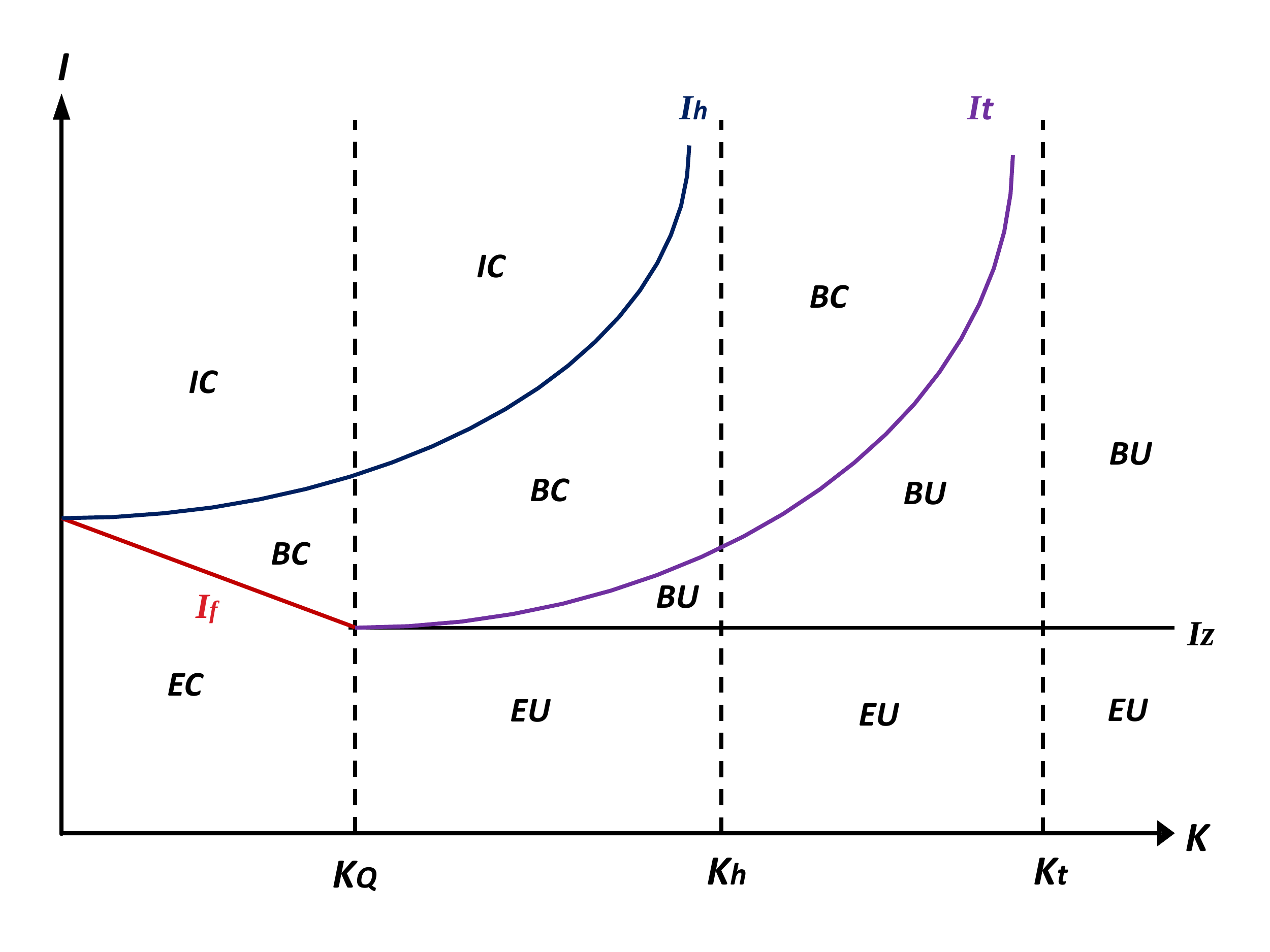} \\
  \caption{Regions}
\label{five regions}
\end{center}
\end{figure}

First we note that when capacity is unlimited, i.e., $K \geq K_{t}$, there is always incentive for the entrant manufacturer to sell in its own domestic market. From Lemma 4.1, the entrant manufacturer sells in the incumbent market in the beginning of the planning horizon if the exchange rate is above the threshold given in (\ref{iz}). In case of limited capacity, any one of the three scenarios (home only, foreign only or split capacity) may occur depending on the exchange rate as shown in Proposition 4.2.

Suppose that capacity is very limited (high-cost case), that is, lower than $K_{Q} $. This situation is more likely to happen when the cost of capacity is too high that the entrant manufacturer cannot fully satisfy the demand in the domestic market alone. In this case, the capacity is fully utilized. As shown in Figure \ref{five regions}, the entrant manufacturer allocates all of its capacity to its own market (EC) if the exchange rate is below $I_{f}$. In this case the low exchange rate prohibit the manufacturer from exporting to the incumbent market. The entrant manufacturer  chooses to allocate the capacity between the two markets (BC) if the exchange rate $I(t)$ falls between $I_{h}\,$ and $I_{f}$. 
If the exchange rate is too high, i.e., $ I(t) > I_{h}$,  the entrant manufacturer will allocate all the limited capacity to the incumbent's market (IC).

Under this scenario ex-ante expected profit function before the start of the selling season can be written as follows:
\begin{equation}
\max_{K}: \Pi_{e} (k) = - u  K + \int_0^{I_{f}}\Pi_e^{EC}A(I)dI +\int_{I_{f}}^{I_{h}}\Pi_e^{BC}A(I)dI+\int_{I_{h}}^{\infty}\Pi_e^{IC}A(I)dI
\label{capacity decision 1}
\end{equation}
Where:
\begin{equation*}
\Pi_e^{EC} = (\alpha_{e} -  C_{e} - K) K \,
\end{equation*}
\begin{equation*}
\Pi_{e}^{BC}(K)=\left(P^{*}_{ee}(K) -C_e\right) \left(\alpha_{e} -  P^{*}_{ee}(K) \right)+\left(I P^{*}_{ei}(K) - C_e - s \right) \left( (1-\phi) \alpha_{i}-  P^{*}_{ei}(K)+ \theta P^{*}_{ii}(K) \right)  \,
\end{equation*}
\begin{equation*}
\Pi_e^{IC} = \frac{[I (2 \alpha_{i}  (1- \phi) + \alpha_{i} \theta  \phi +  C_{i} \theta -2  K) - (C_{e}+s) ( 2 - \theta ^2 ) ] K}{2 -\theta ^2} \ . 
\end{equation*}
To find the optimal capacity, we need to take the derivative of (\ref{capacity decision 1}) with respect to $ K $: 
\begin{equation*}
\frac{d\Pi_{e} (k)}{dK}=-u +\int_0^{\frac{I_{f}}{I_o}}  (\alpha_{e} -  C_{e} - 2 K)  f(\epsilon)d\epsilon + \int_{
\frac{I_f}{I_o}}^{\frac{I_{h}}{I_o}}\lambda f(\epsilon)d\epsilon 
\end{equation*}
\begin{equation}
\hspace{5cm} + \int_{\frac{I_{h}}{I_o}}^\infty[\frac{I (2 \alpha_{i}  (1- \phi) + \alpha_{i} \theta  \phi +  C_{i} \theta -4  K)  }{2 -\theta ^2}  - C_{e}+s] f(\epsilon)d\epsilon =0 \label{capacity decision 1d} 
\end{equation}
where $\lambda$ is the Lagrangian multiplier defined in (5.21). It is straightforward to see from the function above that the second derivative with respect to $K$ is strictly negative for any $K \geq 0$ implying concavity. Hence solution to (\ref{capacity decision 1d}) gives the unique optimal value for the capacity to be built at home in advance of the selling season. Unfortunately, there is no close form solution. However, the optimal capacity can be calculated easily with a simple line search. 

Recall that the condition given in (\ref{capacity decision 1d}) applies when the capacity scarce. This is expected to occur when capacity is very expensive. Suppose that capacity is bigger than $K_{Q}$ then capacity allocation decision according to the following four regions as shown in Figure \ref{five regions} :

(1) if the spot exchange rate is below $I_{z}$, then the manufacturer satisfies all the demand in its own market and stays out of the incumbent market. 

(2) if the spot exchange rate falls between $I_{z}$ and $I_{t}$, then the manufacturer satisfies all demand in both markets. 

(3) if the spot exchange rate falls between $I_{t}$ and $I_{h}$, then the manufacturer splits its limited capacity between the two markets. 

(4) if the spot exchange rate is higher than $I_{h}$, then the manufacturer allocates all the capacity to the incumbent market. 

Consequently, the manufacturer's profit function in this case can be written as:
\begin{equation}
\max_{K}: \Pi_{e} (k) = - u  K + \int_0^{I_{z}}\Pi_e^{EU}A(I)dI+ \int_{I_{z}}^{I_{t}}\Pi_e^{BU}A(I)dI +\int_{I_{t}}^{I_{h}}\Pi_e^{BC}A(I)dI+\int_{I_{h}}^{\infty}\Pi_e^{IC}A(I)dI
\label{capacity decision 2}
\end{equation}
In order to find the optimal capacity, we need to take the derivative of equation (\ref{capacity decision 2}) with respect to $K$ and we get:
\begin{equation}
\frac{d\Pi_{e} (k)}{dK}= -u +  \int_{
\frac{I_f}{I_o}}^{\frac{I_{h}}{I_o}}\lambda f(\epsilon)d\epsilon+
\int_{\frac{I_{h}}{I_o}}^\infty[\frac{I (2 \alpha_{i}  (1- \phi) + \alpha_{i} \theta  \phi +  C_{i} \theta -4  K) }{2 -\theta ^2}  - C_{e}+s] f(\epsilon)d\epsilon =0 \label{capacity decision 2d} 
\end{equation}

Suppose that capacity falls between $K_{h}$ and $K_{t}$ where the firm will always satisfy demand at home market. The capacity allocation decision is based on  the following four regions as shown in Figure \ref{five regions} :

(1) if the spot exchange rate is below $I_{z}$, then the manufacturer satisfies all the demand in its own market and stays out of the incumbent market. 

(2) if the spot exchange rate falls between $I_{z}$ and $I_{t}$, then the manufacturer satisfies all demand in both markets. 

(3) if the spot exchange rate is higher than $I_{h}$, then the manufacturer splits its limited capacity between the two markets. 

Therefore,  the manufacturer's profit function in this case can be written as
\begin{equation}
\max_{K}: \Pi_{e} (k) = - u  K + \int_0^{I_{z}}\Pi_e^{EU}A(I)dI+ \int_{I_{z}}^{I_{t}}\Pi_e^{BU}A(I)dI +\int_{I_{t}}^{I_{h}}\Pi_e^{BC}A(I)dI
\label{capacity decision 3}
\end{equation}
In order to find the optimal capacity, we need to take the derivative of equation (\ref{capacity decision 3}) with respect to $K$ and we can get:
\begin{equation}
\frac{d\Pi_{e} (k)}{dK}= -u +  \int_{
\frac{I_f}{I_o}}^{\frac{I_{h}}{I_o}}\lambda f(\epsilon)d\epsilon \ . \label{capacity decision 3d} 
\end{equation}

Although we cannot find closed form solutions for (\ref{capacity decision 1d}), (\ref{capacity decision 2d}), and (\ref{capacity decision 3d}) we can derive the following conclusions from the property of concavity. \\

\noindent \MakeUppercase{P{\small roposition }} 5.3:  {There exit two capacity cost thresholds $u^{t1}$ and $u^{t2}$ such that $u^{t1}$>$u^{t2}$.  if $u >u^{t1}$, then $K^{^{\ast}}$ will be obtained from (\ref{capacity decision 1d}), if $u_l^{t2}>u_l>u_l^{t1}$, then $K^{^{\ast}}$will be obtained from (\ref{capacity decision 2d}), and if $u^{t2}_l>u_l$, then $K^{^{\ast}}$will be obtained from (\ref{capacity decision 3d}). }

The above proposition shows that when capcity cost is expensive, then the optimal capacity falls between $0$ and $k_{Q}$. However, as capacity cost gets cheaper, then optimal capacity falls between $K_{h}$ and $K_{Q}$, otherwise, the optimal capacity falls between $K_{h}$ and $K_{t}$. \\

\begin{figure}[ht]
\begin{center}
  \includegraphics[scale=0.5]{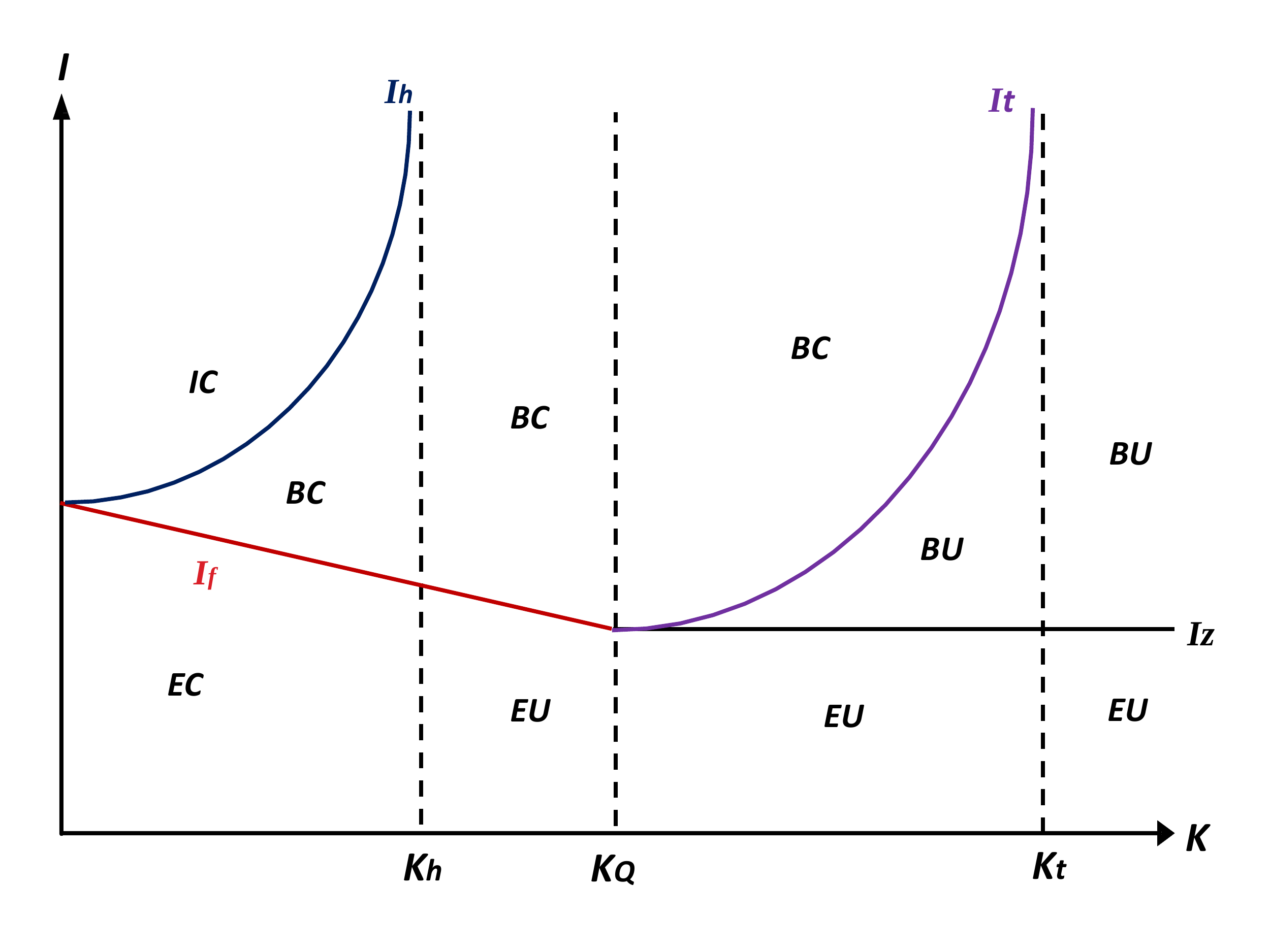} \\
  \caption{Regions II}
\label{five regions II}
\end{center}
\end{figure}






\section{Conclusions and Future Research}
\noindent In the Competition setting, we showed the impact of competition on optimal prices, quantities, and capacity under two scenarios: Exogenous Model and Endogenous Model. Under the Exogenous model, we showed the impact of the foreign manufacturer price on local sales and the firm's decision to enter the foreign
market. As the foreign manufacturer's price increases, it becomes more appealing for the firm to enter the foreign manufacturer, sales at home country decreases, and the optimal capacity increases. Also, as exchange rate volatility increases, optimal capacity decreases.

Under the Endogenous Model, where capacity is ample, it becomes attractive to export to the foreign market when "cross-price", cost per unit for the foreign manufacturer, potential demand at the foreign country, and consumer sensitivity for the foreign price increases. We presented exchange rate thresholds when the firm generate more money than the foreign manufacturer, and when the prices at the foreign country are identical. Also, we showed that high cost per unit for the foreign manufacturer makes the firm the only player at the foreign market. When capacity is ample, sales at the home market stays constant. However, when capacity is limited, the firm must allocate its capacity properly between the two markets to attain optimal profit. We drove exchange rate thresholds where: the firm split capacity between the two markets, sell only to home market, and sell only to foreign market. We showed that an increase in the drift of exchange rate, potential demand, and "cross-price" increase optimal capacity. On the other hand, an increase in consumer price sensitivity, and exchange rate volatility decreases optimal capacity.

In future study, The future work is to make demand uncertain. Also, using different demand functions such as Cobb-Douglas function (see Gallego and Hu (2006)), the CES function (see Varian (1992)), or the Transcendental Logarithmic (translog) function (see Christensen et al. (1973)). In addition to exchange rate, we also aim to integrate the impact of after-tax profit calculations and tariffs into our analysis.

\section*{References}

\begin{hangref}

\item Ahmed A., M. Erkoc, and S. Cho. 2012. Capacity investment, pricing, and production allocation across international markets with exchange rate uncertainty. {\it Asia-Pacific Journal of Operational Research} {\bf 29}(01) 1240008 --1/34.

\item Anderson, K. (February 21, 1997). The Uses and Abuses of Risk Management: How Men Learnt to Bet Against the Gods. Times Literary Supplement

\item Anupindi,  R.,  L. Jiang.  2008.  Capacity  investment under  postponement strategies,  market competition,  and demand  uncertainty. {\it Management Science} {\bf 54}(11) 1876--1890.

\item Aytekin, U., J. R. Birge.  2004. Optimal Investment and Production Across Markets with Stochastic Exchange Rates. Working paper, Northwestern University and The University of Chicago.

\item Bernstein, F., A.  Federgruen. 2004. Dynamic inventory and pricing models for competing retailers. {\it Naval Research Logistics} {\bf 51}(2) 258--274. 

\item Bernstein, F.,  A.  Federgruen. 2005. Decentralized supply chains with competing retailers under demand uncertainty. {\it Management Science} {\bf 51}(1) 18--29.

\item Bitran, G., R. Caldentey.  2003.  An overview of pricing models for revenue management. {\it Manufacturing $\&$ Service Operations Management} {\bf 5}(3) 203--229.

\item Chan, L. M. A.,  Z. J. M. Shen, D. Simchi-Levi, and J. L. Swann.  2004. Coordination of Pricing and Inventory Decisions: A Survey and Classification. In D. Simchi-Levi, S. D. Wu, $\&$ Z. J. M. Shen, (Eds.), Handbook of Quantitative Supply Chain Analysis: Modeling in the E-Business Era 335-392. Kluwer Academic Publishers.

\item Chowdhry,  B.,   J. T. B.  Howe  1999.  Corporate  risk  management  for multinational corporations: Financial and operational hedging policies. {\it European Finance Review} {\bf 2}(2) 229--246.

\item Cohen, M.A., S. Mallik. 1997. Global supply chains: Research and applications. {\it Production and Operations Management} {\bf 6}(2) 193--210.

\item Cohen, M. A., A. Huchzermeier. 1999. Global Supply Chain Management: A survey of research and applications. In S. Tayur, M. Magazine, $\&$ R. Ganeshan (Eds.) Quantitative Models for Supply Chain Management. Kluwer Academic Publishers.

\item Dasu, S., L. Li.  1997. Optimal Operating Policies in the Presence of Exchange Rate Variability. {\it Management Science} {\bf 43}(5), 705--722.

\item Davis, M. 2001. Mathematical of financial markets. Mathematics unlimited. Bjorn
Engquist; and Wilfried Schmid ed. In The Statistical In The Statistical Mechanics of Financial Markets, 3rd edn., pp. 361--380. Berlin: Springer.


\item Ding, Q., L. Dong, and P. Kouvelis.  2007.  On the integration of production and financial hedging decisions in global markets. {\it Operations Research} {\bf 55}(3) 470--489.

\item Dornier, P.  1998. {\it Global Operations and Logistics: Text and Cases.} New York: John Wiley $\&$ Sons.


\item Huchzermeier, A., M. Cohen. 1996. Valuing operational flexibility under exchange rate risk. , {\it Operations Research} {\bf 44}(1) 100-113.

\item Kazaz, B., M. Dada, and H. Moskowitz. 2005. Global production planning under exchange-rate uncertainty. {\it Management Science} {\bf 51}(7) 1101--1119.

\item Kogut, B., N. Kulatilaka. 1994. Operating flexibility, global manufacturing, and option value of a multinational Network.  {\it Management Science} {\bf 51}(7), 123--139.

\item Wen, L., H. Yang, H., D. Bu, L. Diers,  and H. Wang.  2018. Public accounting vs private accounting, career choice of accounting students in China. {\it Journal of Accounting in Emerging Economies}  Vol. 8 Issue: 1, pp.124-140. 

\item Kouvelis, P., M. J. Rosenblatt. 2001. A Mathematical Programming Model to Global Supply Cain Management: Conceptual Approach and Managerial Insights. Working paper, Washington University and Technion

\item Kouvelis, P., K. Axarloglou, and V. Sinha. 2001. Exchange rates and the choice of ownership structure of production facilities. {\it Management Science} {\bf 47}(8) 1063--1080.

\item Li, X., Liao, X.,  Wang, H., \& Wei, L. 2018. Coordinating data pricing in closed-loop data supply chain with data value uncertainty (February 8, 2018).  Available at SSRN:  http://dx.doi.org/10.2139/ssrn.3120590

\item Singh, S.,  H. Wang, and M. Zhu. 2017.  Perceptions of Social Loafing in Groups: Role of Conflict and Emotions. Available at SSRN: http://dx.doi.org/10.2139/ssrn.3132871

\item Singh, S.,  H. Wang, and M. Zhu. 2017.  Perceptions of Social Loafing during the Process of Group Development”. Midwest Academy of Management Conference Proceedings Chicago, Illinois. Working paper. 

\item Lowe, T., R. E. Wendell, and G. Hu. 2002. Screening Location Strategies to Reduce Exchange Rate Risk. {\it European Journal of Operational Research} {\bf 136}(3) 573--590.

\item Mello, A., J. Parsons, and A. Triantis. 1995. An Integrated model of multinational flexibility and financial hedging. {\it Journal of International Economics} {\bf 39}(1-2) 27--51. 

\item O'Brien, T. J. 1996. Global Financial Management. New York: Wiley.

\item Van Mieghem, J.A.,  M. Dada. 1999. Price versus production postponement: Capacity and competition. {\it Management Science} {\bf 45}(12) 1631--1649.

\item Wang, H. 2014. Essays on price competition and strategies: market entry with capacity allocation, channel design and information provision. Open Access Dissertations. 1339.  

https://scholarlyrepository.miami.edu/oa\_dissertations/1339

\item Wang, H., H. Gurnani, and  M. Erkoc. 2016. Entry deterrence of capacitated competition using price and non-price strategies. {\it Production and Operations Management}  {\bf 24}(4) 719 --735.

\end{hangref}

\newpage
\setcounter{page}{1}    
\pagenumbering{arabic}

\end{document}